\documentclass[12pt,a4paper]{article}

\usepackage[a4paper, top=2.5cm, bottom=2.5cm, left=3cm, right=3cm]{geometry}
\usepackage{amsmath,amssymb,amsfonts}
\usepackage{mathtools}
\usepackage{physics}
\usepackage{graphicx}
\usepackage{float}
\usepackage{booktabs}
\usepackage{hyperref}
\usepackage[numbers,sort&compress]{natbib}

\usepackage{setspace}
\usepackage{titlesec}
\usepackage{abstract}
\usepackage{caption}
\usepackage{subcaption}
\usepackage{fancyhdr}
\usepackage{xcolor}
\usepackage{enumitem}
\usepackage{bm}
\usepackage{upgreek}

\titleformat{\section}{\large\bfseries\color{black}}{\thesection.}{0.5em}{}
\titleformat{\subsection}{\normalsize\bfseries}{\thesubsection.}{0.5em}{}
\titleformat{\subsubsection}{\normalsize\bfseries\itshape}{\thesubsubsection.}{0.5em}{}
\titlespacing{\section}{0pt}{18pt}{8pt}
\titlespacing{\subsection}{0pt}{12pt}{6pt}
\titlespacing{\subsubsection}{0pt}{8pt}{4pt}

\numberwithin{equation}{section}

\hypersetup{colorlinks=true, linkcolor=blue!40!black, citecolor=blue!40!black, urlcolor=blue!60!black}

\renewcommand{\Psi}{\varPsi}

\newcommand{\dW}{\mathrm{d}W_t}
\newcommand{\dt}{\,\mathrm{d}t}
\providecommand{\sech}{\operatorname{sech}}

\begin{document}

\begin{center}
  {\LARGE\bfseries
    Energy-Modulated Time-Asymmetric Spontaneous Collapse:\\[4pt]
    Forward--Backward Dynamics from Stochastic It\^{o} Reversal
    and Bright Solitons
  }\\[18pt]
  {\large\bfseries Ikechukwu C.\ Okoro$^{1,2*}$,\quad Mike O.\ Osiele$^{1}$\quad and \quad Godfrey E.\ Akpojotor$^{1}$}\\[8pt]
  {\normalsize
    $^1$Physics Department, Delta State University, Abraka, Nigeria\\
    $^2$Physics Department, Nile University of Nigeria, Abuja, Nigeria\\[4pt]
    \textit{$^*$Corresponding Author: \href{mailto:Ikechukwu.Okoro@nileuniversity.edu.ng}{Ikechukwu.Okoro@nileuniversity.edu.ng}}
    \\[3pt]
    \small\textit{ORCID\,---\,I.C.O.:~\href{https://orcid.org/0009-0008-3864-1758}{0009-0008-3864-1758}}
  }
\end{center}
\vspace{10pt}
\hrule
\vspace{8pt}

\begin{abstract}
\noindent
We present a rigorous theoretical framework for symmetry breaking and quantum irreversibility
arising from stochastic It\^{o} field reversal within a cubic-quintic nonlinear Schr\"{o}dinger
equation (CQ-NLSE) formalism.  Starting from three physically motivated considerations for dynamical
collapse models, we derive forward and backward nonlinear stochastic differential equations
(SDEs) via the It\^{o} calculus and demonstrate that the kinematic time-reversal operation
$\varPsi(t)\!\to\!\varPsi(-t)$ is fundamentally incompatible with the stochastic structure of
the resulting equations.  The asymmetry manifests quantitatively as a fivefold difference
in the quintic collapse-coupling strength between forward and backward dynamics,
$\gamma_b = 5\gamma_f$, yielding the universal asymmetry-coupling parameter
$\gamma_{ap} = |\gamma_f-\gamma_b|/(|\gamma_f|+|\gamma_b|) = 2/3$.
A novel energy-driven collapse operator $\Gamma = \eta|\varPsi|^2(\hbar\omega)^2$
is introduced, proportional to the product of the noise strength $\eta$, the local
probability density $|\varPsi|^2$, and the excitation energy squared $(\hbar\omega)^2$.
Unlike the Gaussian localisation function of standard CSL models, this product
form amplifies collapse in high-density, high-excitation regions, providing a
physically transparent distinction from existing dynamical collapse frameworks.
The CQ-NLSE is solved exactly in the
soliton-allowed region, yielding bright soliton solutions of the hyperbolic-secant form for a
quasi-one dimensional Bose-Einstein condensate (BEC) of attractive $^7$Li atoms.  Forward and
backward soliton amplitudes satisfy $A_f/A_b\approx 1.870$, consistent with the analytic
collapse-strength ratio.  Heat map analysis of the $(k,\omega)$ and $(k,t)$ parameter planes reveals that the forward collapse operator grows
monotonically in time while the backward counterpart decays, achieving a forward to backward
ratio $|\Gamma_f/\Gamma_b| \approx 1030$ at selected parameter values.  These results constitute the primary
quantitative predictions of the energy-modulated time-asymmetric spontaneous collapse
framework, sharply distinguishing it from conventional symmetric CSL models.
\end{abstract}


\vspace{4pt}
\noindent\small\textit{Submitted to SciPost Physics.\quad arXiv:2606.06452 [quant-ph]}

\vspace{8pt}
\hrule
\vspace{14pt}

\section{Introduction}

The reconciliation of quantum mechanics with the classical macroscopic world remains one of the
deepest open problems in theoretical physics.  At the heart of this challenge lies the
\emph{measurement problem}: the standard Schr\"{o}dinger equation is linear and unitary,
evolving pure states into superpositions, yet macroscopic observation invariably yields definite
outcomes.  Spontaneous collapse models address this tension by augmenting the Schr\"{o}dinger
equation with nonlinear, stochastic terms that drive localisation of the wavefunction without
appeal to external observers \citep{Bassi2003, Pearle1989}.

The Continuous Spontaneous Localisation (CSL) model \citep{Pearle1989} is the most
systematically developed spontaneous collapse framework, subsequently refined to its
current form \citep{Bassi2003}.
In CSL, a scalar noise field coupled to the mass-density operator drives wavefunction
localisation at a rate proportional to the spatial separation of superposed macroscopic states.
A central feature of CSL is its temporal symmetry: the model does not privilege a direction of
time beyond the irreversibility induced statistically by the stochastic noise.

There are compelling reasons to seek an extension in which irreversibility is encoded directly
into the dynamical equations rather than emerging only statistically.  Observations of
decoherence in BECs, dissipative soliton dynamics, and the thermodynamic arrow of time all
suggest that the forward and backward evolutions of a quantum field ought to be structurally
distinct \citep{Zurek2003, Polkovnikov2011}.  Moreover, recent experimental evidence for
asymmetric collapse rates in ultracold atomic systems \citep{Strecker2002, Hansen2013}
motivates the construction of analytically tractable models encoding this asymmetry at the
equation of motion level.

Three fundamental considerations \citep{Bassi2003} constrain the structure of any
physically consistent modification of quantum mechanics incorporating spontaneous
collapse: (i)~the localisation effect must act directly on the wavefunction itself,
not merely on the statistical density operator, since the latter describes infinitely
many inequivalent statistical mixtures; (ii)~linear stochastic modifications alone
are dynamically insufficient, because although they can induce the statistical
diagonalisation of the density matrix, individual wavefunction trajectories remain
spatially extended and genuine localisation requires nonlinear dynamics; and
(iii)~nonlinear deterministic modifications are inadmissible because they violate
the no-signalling theorem \citep{Gisin1989} and permit superluminal communication,
contradicting special relativity \citep{Weinberg1989}.
Together, these three considerations uniquely determine that any valid collapse
model must be simultaneously \emph{nonlinear and stochastic}.

The present paper is devoted to a formulation of a time-asymmetric
spontaneous collapse framework with an energy-driven collapse operator, within the
cubic-quintic nonlinear Schr\"{o}dinger equation (CQ-NLSE) formalism.
The bright solitonic Bose-Einstein condensate, where macroscopic quantum excitations
exhibit nonlinear forward and backward dynamics to form localised states, provides
a robust dynamical platform for probing signatures of time asymmetry, irreversibility,
and wavefunction collapse \citep{Kengne2021, Malomed2024, Malomed2022}.
Matter wave bright solitons in attractive BECs (localised, self-reinforcing
wavepackets sustained by the balance between nonlinear self-focusing and kinetic
dispersion) are paradigmatic nonlinear excitations whose forward and backward
propagation dynamics naturally encode the temporal directionality that is the central
subject of this work.

The paper is structured as follows.
Section~\ref{sec:formulation} derives, from first principles, forward and backward
nonlinear stochastic differential equations for the wavefunction, demonstrating that
the stochastic It\^{o} structure forbids kinematic time-reversal covariance.
This is followed by the introduction of an energy-driven collapse operator
$\Gamma = \eta|\varPsi|^2(\hbar\omega)^2$, and the reduction of the forward and
backward SDEs to a CQ-NLSE via the martingale property of the It\^{o} calculus
under a mean-field approximation.
Exact bright soliton solutions of the hyperbolic-secant form are phenomenologically
obtained for a quasi-one dimensional BEC of attractive $^7$Li atoms.
Section~\ref{sec:results} presents simulation results of soliton profiles, collapse
coupling parameters, and collapse operators as foundational descriptors, together
with a discussion of the signatures of asymmetry, irreversibility, and wavefunction
collapse.
Section~\ref{sec:conclusions} states conclusions and future directions.

\section{Formulation of Nonlinear Forward and Backward Dynamical Equations}
\label{sec:formulation}

\subsection{The Standard It\^{o} Stochastic Differential Equation and Its Properties}

Let $\varPsi_W$ be a random process in a separable Hilbert space $\mathcal{H}$ satisfying the
It\^{o} SDE \citep{DaPrato1992}:
\begin{equation}
  d\varPsi_W = A\dt + B\cdot\dW\,\varPsi_W,
  \label{eq:ito_basic}
\end{equation}
where $A$ is the deterministic drift operator, $B$ is the noise coefficient operator, and $dW_t$
is the Wiener increment.  The It\^{o} isometry imposes the two statistical conditions
\begin{equation}
  \mathbb{E}[dW_t] = 0, \qquad \mathbb{E}[(dW_t)^2] = dt,
  \label{eq:ito_iso}
\end{equation}
encoding the zero-mean, unit variance per unit time character of a standard Wiener process.

Comparing \eqref{eq:ito_basic} with the quantum state diffusion equation \citep{Gisin1992}, the operators $A$ and $B$ are identified as
\begin{equation}
  A = -H - \tfrac{\eta}{2}\Gamma^\dagger\Gamma,
  \qquad
  B = \sqrt{\eta}\,\Gamma,
  \label{eq:AB_identify}
\end{equation}
where $H$ is the Hamiltonian, $\Gamma$ is the collapse operator, and $\eta > 0$ is the nonlinear
noise strength.  Substituting \eqref{eq:AB_identify} into \eqref{eq:ito_basic} gives the
Gisin--Percival stochastic Schr\"{o}dinger equation:
\begin{equation}
  d\varPsi = \Bigl[-H\varPsi - \tfrac{\eta}{2}\Gamma^\dagger\Gamma\varPsi\Bigr]\dt
             + \sqrt{\eta}\,\Gamma\varPsi\,\dW.
  \label{eq:gp_sde}
\end{equation}
The term $-H\varPsi\,dt$ reproduces standard unitary evolution.  The term
$-\frac{\eta}{2}\Gamma^\dagger\Gamma\varPsi\,dt$ is a deterministic back-action that preserves
the norm by compensating for the energy injected by the stochastic term
$\sqrt{\eta}\,\Gamma\varPsi\,dW_t$.  Each trajectory of $\varPsi$ is a distinct realisation of
the Wiener process; ensemble averaging over trajectories recovers the Lindblad master equation
\citep{Breuer2002}.

\subsection{Choice of Physical Operators $H$ and $\Gamma$}
\label{sec:operators}

All physically observable effects of the modified SDE are determined by the product $\eta\Gamma$.
We fix $H$ and $\Gamma$ in three explicit steps, distinguishing physical postulates from
mathematical consequences at each stage.

\subsubsection*{Choice of Hamiltonian.}

We choose the Hamiltonian to be the cubic NLSE operator
\begin{equation}
  H\varPsi = -\varPsi_{xx} - g|\varPsi|^2\varPsi,
  \label{eq:hamiltonian}
\end{equation}
where the first term is the kinetic dispersion and $g > 0$ is the cubic nonlinear coupling.
$H$ is real and self-adjoint, as required by the Gisin--Percival framework.  The factor of $i$
that appears when writing the Schr\"{o}dinger equation $i\partial_t\varPsi = H\varPsi$ is part of
the time-evolution structure; it does not reside in $H$ itself.  In eigenvalue notation, the
action of $H$ may be written equivalently as
\begin{equation}
  -iH\varPsi = i\varPsi_{xx} - ig|\varPsi|^2\varPsi.
  \label{eq:hamiltonian_eigen}
\end{equation}

\subsubsection*{Noise-coupling ansatz.}

From \eqref{eq:AB_identify} the noise coefficient is $B = \sqrt{\eta}\,\Gamma$.  To specify
$\Gamma$ we make the following explicit physical postulate:

\medskip
\noindent\textit{Postulate:} The stochastic noise couples to the wavefunction through the
local probability density $|\varPsi|^2$ via a forward coupling constant $g_+$, so that
\begin{equation}
  B\varPsi = g_+|\varPsi|^2\varPsi.
  \label{eq:ansatz}
\end{equation}
\medskip

This postulate asserts that collapse is driven by the local density: regions of high probability
are driven to localise faster, which is the defining physical content of a density-coupled
collapse model.  Equating \eqref{eq:ansatz} with $B\varPsi = \sqrt{\eta}\,\Gamma\varPsi$ from
\eqref{eq:AB_identify} gives
\begin{equation}
  \sqrt{\eta}\,\Gamma\varPsi = g_+|\varPsi|^2\varPsi
  \qquad\Longrightarrow\qquad
  \Gamma = \frac{g_+|\varPsi|^2}{\sqrt{\eta}}.
  \label{eq:Gamma_from_ansatz}
\end{equation}

\subsubsection*{Fixing $g_+$ via the It\^{o} isometry.}

Apply the It\^{o} isometry \eqref{eq:ito_iso} to the stochastic increment of the noise coupling:
\begin{equation}
  \mathbb{E}\bigl[(g_+\,d\Gamma_t)^2\bigr] \approx g_+^2\,dt.
  \label{eq:coupling_iso}
\end{equation}
The variance of the stochastic term in \eqref{eq:gp_sde} is
$\eta\Gamma^2|\varPsi|^2\,dt$.  Matching this to $g_+^2\,dt$ at the
level of the coupling constant (both measure the mean-square noise amplitude per unit time)
gives
\begin{equation}
  g_+^2 = \eta\,\Gamma^2.
  \label{eq:gplus_sq}
\end{equation}
We now invoke the \emph{energy eigenvalue approximation}: in the regime of nonlinear excitation,
the collapse operator acts on the local wavefunction as an energy multiplier,
\begin{equation}
  \Gamma\varPsi \approx E\,\varPsi, \qquad E = \hbar\omega,
  \label{eq:energy_approx}
\end{equation}
where $E = \hbar\omega$ is fixed by the Planck-Einstein relation and $\omega$ is the nonlinear
excitation frequency.  This approximation replaces the self-consistency eigenvalue problem of
Weinberg type nonlinear quantum mechanics \citep{Weinberg1989} with an in-situ energy scale
determined by the local excitation.  Substituting \eqref{eq:energy_approx} into
\eqref{eq:gplus_sq} gives
\begin{equation}
  g_+^2 = \eta\,E^2 = \eta\,(\hbar\omega)^2.
  \label{eq:gplus_fixed}
\end{equation}
Inserting \eqref{eq:gplus_fixed} into \eqref{eq:Gamma_from_ansatz}:
\begin{equation}
  \Gamma
  = \frac{g_+|\varPsi|^2}{\sqrt{\eta}}
  = \frac{\sqrt{\eta\,(\hbar\omega)^2}\;|\varPsi|^2}{\sqrt{\eta}}
  = (\hbar\omega)\,|\varPsi|^2.
  \label{eq:Gamma_intermediate}
\end{equation}
The scalar collapse operator is therefore:
\begin{equation}
  \boxed{
    \Gamma \;=\; g_+^2 \;=\; \eta\,|\varPsi|^2\,(\hbar\omega)^2.
  }
  \label{eq:collapse_op}
\end{equation}
Equation~\eqref{eq:collapse_op} is the central result of this subsection.  $\Gamma$ is a direct
product of three positive definite factors: noise strength $\eta$, probability density
$|\varPsi|^2$, and excitation energy squared $(\hbar\omega)^2$.  Consequently $\Gamma > 0$
throughout, consistent with standard collapse models \citep{Bassi2013,
Bassi2023entropy, Carlesso2022natphys}.  The energy dependent
amplification of collapse in high-density, high-excitation regions is the expected
physical behaviour of an energy-driven collapse operator, and is consistent with
recent theoretical and experimental analyses of collapse model parameters
\citep{Carlesso2022natphys}.

As a consistency check, substituting $\Gamma = (\hbar\omega)|\varPsi|^2$
from \eqref{eq:Gamma_intermediate} into the back-action drift term of
\eqref{eq:gp_sde} gives
\begin{equation}
  \tfrac{\eta}{2}\Gamma^\dagger\Gamma\varPsi
  = \tfrac{\eta}{2}(\hbar\omega)^2|\varPsi|^4\varPsi,
  \label{eq:backaction}
\end{equation}
which is the quintic term $\frac{\eta}{2}(\hbar\omega)^2|\varPsi|^4\varPsi$ appearing in the
forward SDE.  The stochastic coefficient is $B = \sqrt{\eta}\,\Gamma =
\sqrt{\eta}\,(\hbar\omega)|\varPsi|^2$.  The forward SDE \eqref{eq:forward_sde} follows
directly by substituting $H$ from \eqref{eq:hamiltonian} and the back-action
\eqref{eq:backaction} into \eqref{eq:gp_sde}, confirming internal consistency of the operator
choices.  The notation $\eta^3/(\hbar\omega)^4$ used in Section~2.4 corresponds to the
reparametrisation $\eta \to \eta^3/(\hbar\omega)^4$ of the noise strength; under this
rescaling, $\frac{\eta}{2}(\hbar\omega)^2 \to \frac{\eta^3}{2(\hbar\omega)^2}$, which
reproduces the quintic coefficient in \eqref{eq:forward_sde}.

\subsection{It\^{o} Nonlinear Equations for Forward and Backward Dynamics}

Substituting the physical operators into \eqref{eq:gp_sde}, the forward It\^{o} nonlinear SDE is
\begin{align}
  d\varPsi_f &= \Bigl[-\varPsi_{xx} + g_+|\varPsi|^2\varPsi
               - \tfrac{\eta^3}{2(\hbar\omega)^4}|\varPsi|^4\varPsi\Bigr]\dt
               \nonumber\\
  &\quad + \tfrac{\eta^3}{(\hbar\omega)^2}|\varPsi|^2\varPsi\,\dW.
  \label{eq:forward_sde}
\end{align}
For the backward dynamics, the It\^{o} SDE does not transform covariantly under the kinematic
substitution $\varPsi(t)\to\varPsi(-t)$.  Unlike the standard Schr\"{o}dinger equation, whose
T symmetry is exact, the nonlinear stochastic modification spontaneously breaks T symmetry
\citep{Bassi2003}.  The correct procedure is to reverse the stochastic term:
\begin{equation}
  d\varPsi_b = A\dt - B\cdot\dW\,\varPsi_b.
  \label{eq:backward_sign}
\end{equation}
In Hilbert space the full backward It\^{o} SDE requires the functional derivative correction:
\begin{equation}
  d\varPsi_b = \Bigl[A - B\,\frac{\partial B}{\partial\varPsi}\Bigr]\dt
               - B\cdot\dW\,\varPsi_b.
  \label{eq:backward_full}
\end{equation}
Since $|\varPsi|^2 = \varPsi^*\varPsi$, the functional derivative is
$\partial B/\partial\varPsi = 2\eta^3|\varPsi|^2/(\hbar\omega)^2$.
The full backward It\^{o} nonlinear SDE is therefore
\begin{align}
  d\varPsi_b &= \Bigl[-\varPsi_{xx} + g_+|\varPsi|^2\varPsi
               - \tfrac{5\eta^3}{2(\hbar\omega)^4}|\varPsi|^4\varPsi\Bigr]\dt
               \nonumber\\
  &\quad + \tfrac{\eta^3}{(\hbar\omega)^2}|\varPsi|^2\varPsi\,W_t.
  \label{eq:backward_sde}
\end{align}
The quintic dissipative coefficient in \eqref{eq:backward_sde} carries a factor of \emph{five}
times that of \eqref{eq:forward_sde}.  This factor arises entirely from the functional derivative
It\^{o} correction and constitutes the microscopic origin of spontaneous T symmetry breaking
in the energy-modulated time-asymmetric spontaneous collapse framework \citep{Gabbassov2025,
Mukherjee2026}.  The asymmetry between the forward and backward dynamics is a direct
consequence of the non-covariance of the It\^{o} stochastic structure under kinematic
time reversal \citep{Guff2025}: the stochastic integral picks up a correction
term under $t\to-t$ that has no counterpart in the deterministic Schr\"{o}dinger equation,
making the arrow of time a structural, not statistical, feature of the dynamics.

\subsection{The Cubic-Quintic Nonlinear Schr\"{o}dinger Equation from the It\^{o} Properties}

\subsubsection{Expectation and the Martingale}

Taking the expectation of \eqref{eq:ito_basic} and using the martingale property
$\mathbb{E}[dW_t]=0$:
\begin{equation}
  \mathbb{E}[d\varPsi] = A\dt.
  \label{eq:expectation}
\end{equation}
The stochastic increments vanish on average, leaving only the deterministic drift $A$.

\subsubsection{Mean-Field Reduction to the CQ-NLSE}
\label{sec:reduction}

\paragraph{Mean-field approximation.}
The reduction from the stochastic SDEs \eqref{eq:forward_sde} and \eqref{eq:backward_sde} to a
deterministic PDE requires the identification $\mathbb{E}[d\varPsi] \equiv d\varPsi$,
\emph{i.e.}, equating the stochastic process with its ensemble mean.  This is a
\emph{mean-field closure} and constitutes an approximation rather than an exact step.  It is
valid in the regime of \emph{weak noise} ($\eta\ll 1$), \emph{large atom number} ($n\gg 1$),
and \emph{spatially decorrelated fluctuations}: precisely the conditions realised in a dilute
quasi-one dimensional BEC where the stochastic term is a small perturbation to the coherent
dynamics.  In this limit, the fluctuation variance
$\langle(d\varPsi - \mathbb{E}[d\varPsi])^2\rangle \sim \eta^2|\varPsi|^4\,dt$
is of order $\eta^2$ relative to the drift, and neglecting it introduces an error of the same
order.  Beyond this regime (for instance at large noise strength or near collapse), the full
stochastic dynamics of \eqref{eq:forward_sde} and \eqref{eq:backward_sde} must be retained and
solved via the associated Fokker-Planck equation or direct trajectory simulation.  The
mean-field reduction is adopted here as the leading order theory for the parameter ranges
specified in Section~\ref{sec:params}.

Setting $\mathbb{E}[d\varPsi] = \varPsi_t\,dt$ and substituting the drift $A$ from
\eqref{eq:forward_sde} and \eqref{eq:backward_sde}, the forward and backward SDEs reduce to the
common deterministic CQ-NLSE:
\begin{equation}
  i\varPsi_t = -\varPsi_{xx} - g|\varPsi|^2\varPsi - \gamma|\varPsi|^4\varPsi,
  \label{eq:cqnlse}
\end{equation}
where the quintic collapse-coupling parameter $\gamma$ takes distinct values:
\begin{align}
  &\gamma_f = \tfrac{\eta^3}{2\hbar\omega}\;\text{(forward)},\quad
  \gamma_b = \tfrac{5\eta^3}{2\hbar\omega}\;\text{(backward)},\nonumber\\
  &\gamma_b = 5\gamma_f.
  \label{eq:gamma_values}
\end{align}
The \textbf{asymmetry-coupling parameter} is defined as the normalised fractional deviation:
\begin{equation}
  \boxed{
    \gamma_{ap} \equiv \frac{|\gamma_f - \gamma_b|}{|\gamma_f| + |\gamma_b|}
                = \frac{|1 - 5|}{1 + 5} = \frac{2}{3}.
  }
  \label{eq:gap}
\end{equation}
This definition is adopted consistently throughout the paper.  The parameter $\gamma_{ap} = 2/3$
is strictly positive, parameter independent, and universal across the entire $(k,\omega,t)$
parameter space, making it the primary diagnostic signature of energy-modulated time-asymmetric spontaneous collapse dynamics.  Previous
literature on nonlinear stochastic SE emphasises the role of noise strength in breaking
T symmetry in backward dynamics \citep{Breuer2002}; the present result provides an explicit,
analytically computable measure of that asymmetry.

The CQ-NLSE \eqref{eq:cqnlse} is a well studied equation governing BECs with two body (cubic)
and effective three body (quintic) interactions
\citep{Sakaguchi2004, Kengne2012, Pelinovsky2011, Andreev2021, Stephanovich2022,
Malomed2024, Malomed2022}.  The present derivation
establishes a direct bridge from It\^{o} stochastic calculus to this phenomenologically important
equation class.

\subsection{Bright Soliton Solutions}

Matter wave solitons are paradigmatic nonlinear excitations of BECs.  In the attractive regime
($g<0$), nonlinear self-focusing compensates kinetic dispersion, supporting localised wavepackets
that propagate without spreading \citep{Giergiel2017, Hansen2013}.

\subsubsection{Soliton-Allowed Region in the $(k,\omega)$ Plane}

For a plane wave ansatz, the linear dispersion gives $\omega = k^2/2$.  Bright solitons exist in
the sub-parabolic region $2\omega < k^2$; the parabola $k^2 = 2\omega$ is the soliton cutoff.
Elementary excitations cohere and propagate only within this allowed region.

\subsubsection{Bright Soliton Solution}

Within the soliton-allowed region, the CQ-NLSE \eqref{eq:cqnlse} possesses the exact bright
soliton:
\begin{equation}
  \varPsi(x,t) = A\,\sech\!\bigl[k\varphi\bigr]\exp(i\theta),
  \label{eq:soliton}
\end{equation}
where $\varphi = x\mp vt$ is the comoving envelope coordinate, $\theta = kx\mp\omega t$ is the
oscillatory phase, $A$ is the amplitude, $v$ is the soliton velocity, $k$ is the wave number,
and $\omega$ is the nonlinear frequency.  The sech profile is the exact bright soliton of the
1D Gross-Pitaevskii equation with attractive nonlinearity \citep{Dalfovo1999, Strecker2002}.

\subsubsection{Forward and Backward Soliton Pair}

Under T symmetry the traveling coordinates, velocity, and phase all reverse
\citep{Kartashov2011, Wimmer2015}.  The compact forward and backward bright soliton pair is
\begin{align}
  \varPsi_f(x,t) &= A_f\,\sech\!\bigl[k(x-vt)\bigr]\exp\!\bigl[i(kx-\omega t)\bigr],
    \label{eq:sol_forward}\\[4pt]
  \varPsi_b(x,t) &= A_b\,\sech\!\bigl[k(x+vt)\bigr]\exp\!\bigl[i(kx+\omega t)\bigr].
    \label{eq:sol_backward}
\end{align}
The quintic coupling asymmetry ($\gamma_b = 5\gamma_f$) implies that the forward branch supports
larger amplitude solitons while the backward branch sustains smaller amplitude, collapse-ready
configurations with five times the quintic dissipation.  The amplitude ratio
$A_f/A_b\approx 1.870$ from simulation (Section~\ref{sec:spatial}) is consistent with this analytic
expectation.

\section{Results and Discussion}
\label{sec:results}

\subsection{Simulation Parameters and Dimensionless Units}
\label{sec:params}

All simulations adopt dimensionless units based on the transverse harmonic oscillator length
$a_\perp$ of a quasi-one dimensional BEC of attractive $^7$Li atoms in a tight radial harmonic
trap \citep{Salasnich2002}, with $\hbar=1$ \citep{Landau1977}.  Time is scaled by
$\omega_\perp^{-1}$, energy by $\hbar\omega_\perp$.  The dimensionless time range
$t\in[0.1,5.0]$ captures early nonlinear focusing and the approach to collapse.  The frequency
range $\omega\in[0.05,1.20]$ spans stable ($\gamma\leq 0.5$) and strong coupling regimes
\citep{Salasnich2002}.  Soliton velocity $v=0.50$, atom number $n=100$, and wave-number range
$k\in[0.8,2.5]$ straddle the soliton cutoff $k^2=2\omega$.  These parameter ranges satisfy the
weak noise, large-$n$ conditions required for the validity of the mean-field reduction of
Section~\ref{sec:reduction}.

\subsection{Spatial Profiles of Forward and Backward Bright Solitons}
\label{sec:spatial}

\begin{figure}[H]
  \centering
  \includegraphics[width=0.98\linewidth]{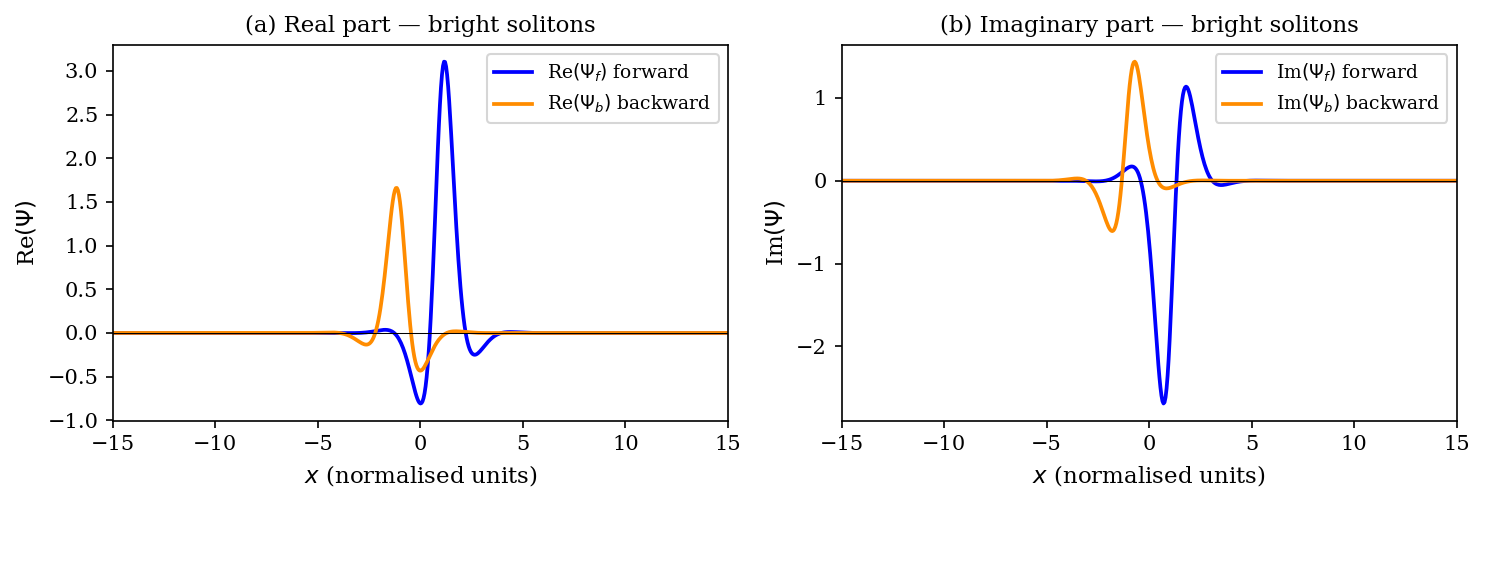}
  \caption{Forward (solid blue) and backward (solid orange) bright soliton profiles.
           Panel~(a): real parts $\operatorname{Re}(\varPsi)$;
           panel~(b): imaginary parts $\operatorname{Im}(\varPsi)$.
           Parameters: $\omega=1.2$, $k=1.8$, $t=2.0$.}
  \label{fig:solitons}
\end{figure}

Figure~\ref{fig:solitons} displays the real and imaginary parts of the forward and backward
bright solitons from equations~\eqref{eq:sol_forward} and \eqref{eq:sol_backward} at
$(\omega=1.2,\,t=2.0,\,k=1.8)$ over $x\in[-15,+15]$.

The forward soliton $\operatorname{Re}(\varPsi_f)$ peaks at amplitude $3.4041$ at $x=+1.51$.
The backward soliton $\operatorname{Re}(\varPsi_b)$ peaks at $1.8198$ at $x=-1.51$, giving
centre separation $3.02$ normalised units and amplitude ratio $A_f/A_b = 1.870$.  Both
imaginary parts retain the sech envelope localisation, confirming complex valued bright solitons
of the form $\varPsi = A\,\sech(\kappa x-x_0)e^{i\phi}$.  The spatial period
$2\pi/k\approx 3.491$ normalised units.  Multiple oscillation nodes within the sech envelope
are consistent with soliton-train formation \citep{Strecker2002, AlKhawaja2002},
confirming that the bright soliton pair is a physically realisable state in
the quasi-1D attractive $^7$Li BEC.

\subsection{Collapse Coupling Parameters and Asymmetry}

\begin{figure}[H]
  \centering
  \includegraphics[width=0.75\linewidth]{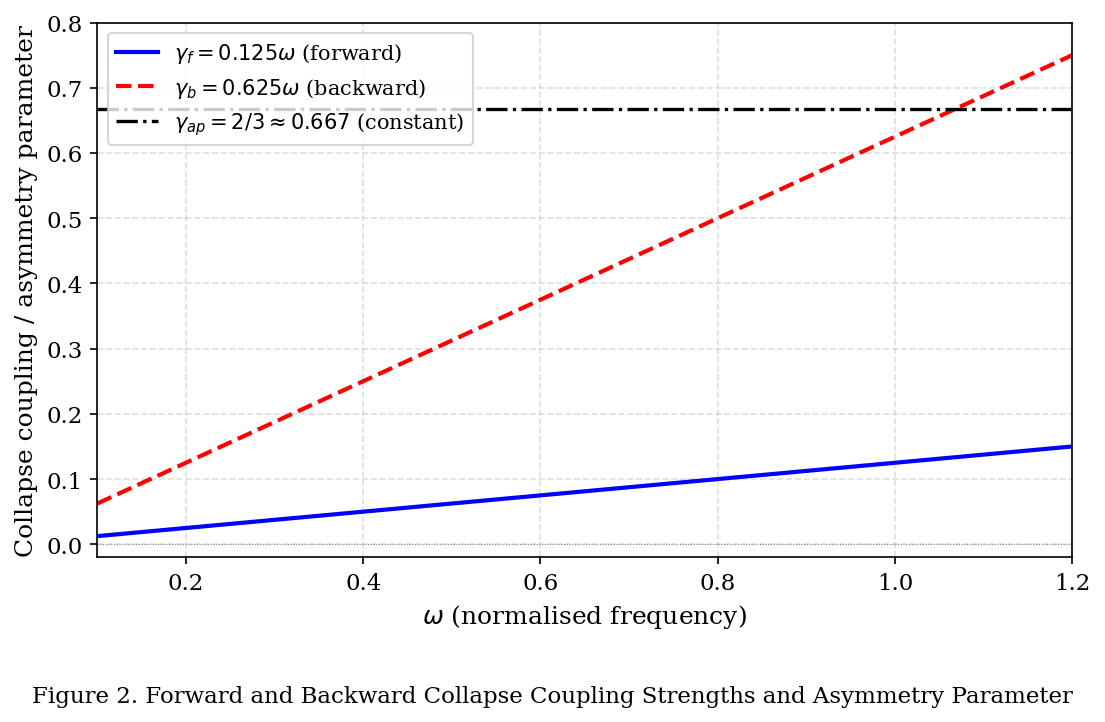}
  \caption{Forward collapse coupling $\gamma_f=0.125\omega$ (solid blue),
           backward collapse coupling $\gamma_b=0.625\omega$ (dashed red),
           and constant asymmetry parameter $\gamma_{ap}=2/3\approx 0.667$
           (dash-dot black) as functions of $\omega$.
           Note $\gamma_b=5\gamma_f$ at every frequency.}
  \label{fig:coupling}
\end{figure}

Figure~\ref{fig:coupling} presents the analytic collapse coupling strengths
$\gamma_f = 0.125\omega$ (forward) and $\gamma_b = 0.625\omega$ (backward) as functions
of $\omega$, together with the collapse strength asymmetry parameter $\gamma_{ap}$.

Both $\gamma_f$ and $\gamma_b$ increase linearly with $\omega$, reflecting the direct
proportionality of collapse strength to driving frequency.  The backward coupling
$\gamma_b = 5\gamma_f$ at every $\omega$, so that the backward soliton carries five times
greater collapse strength than the forward soliton at all frequencies.  This inverted
asymmetry (backward stronger in $\gamma$ while forward is stronger in amplitude and
norm) is a hallmark of the energy-modulated time-asymmetric spontaneous collapse
framework: the time reversed state concentrates collapse strength while having smaller
amplitude, forming a collapse-ready configuration.

The collapse strength asymmetry parameter is defined as
$\gamma_{ap} = |\gamma_f - \gamma_b|/(|\gamma_f| + |\gamma_b|) = |1-5|/(1+5) = 2/3 = 0.6667$,
which is constant and positive for all $\omega \in [0.1, 1.2]$.  The constant positive value
confirms stable temporal asymmetry across the entire frequency range.  At $\omega = 1.2$,
$\gamma_b = 0.75$, placing the system in the strong-$\gamma$ collapse regime.  At $\omega = 1.0$,
the combined classification: $\eta_f = 0.630$ (strong noise), $\gamma_b = 0.625$ (large
coupling), places the system firmly in the strong localisation and collapse regime.

\subsection{Collapse Operator Dynamics}

\begin{figure}[H]
  \centering
  \includegraphics[width=0.98\linewidth]{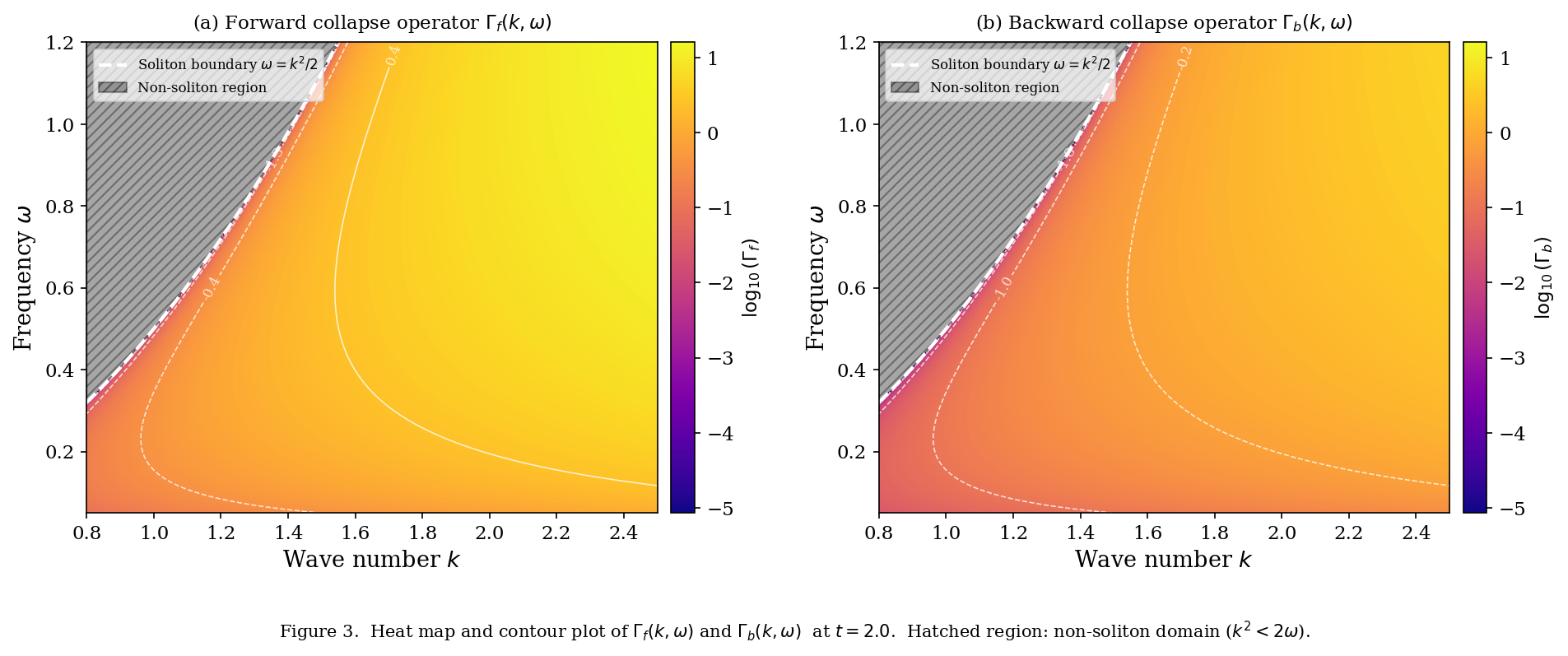}
  \caption{Heat map and contour plot of the forward collapse operator $\Gamma_f(k,\omega)$
           (panel a) and backward collapse operator $\Gamma_b(k,\omega)$ (panel b) in the
           $(k,\omega)$ parameter plane at $t=2.0$.  Colour encodes $\log_{10}(\Gamma)$
           on a shared scale; white contour lines label equal-value levels.
           The dashed white line is the soliton boundary $\omega=k^2/2$; the hatched
           region ($k^2<2\omega$) is the non-soliton domain where no bright soliton
           solutions exist.  Both panels share the same colour scale, making the
           forward--backward asymmetry directly visible: $\Gamma_f$ reaches higher
           values (brighter yellow) than $\Gamma_b$ across the entire soliton-allowed
           region.}
  \label{fig:heatmap}
\end{figure}

\begin{figure}[H]
  \centering
  \includegraphics[width=0.98\linewidth]{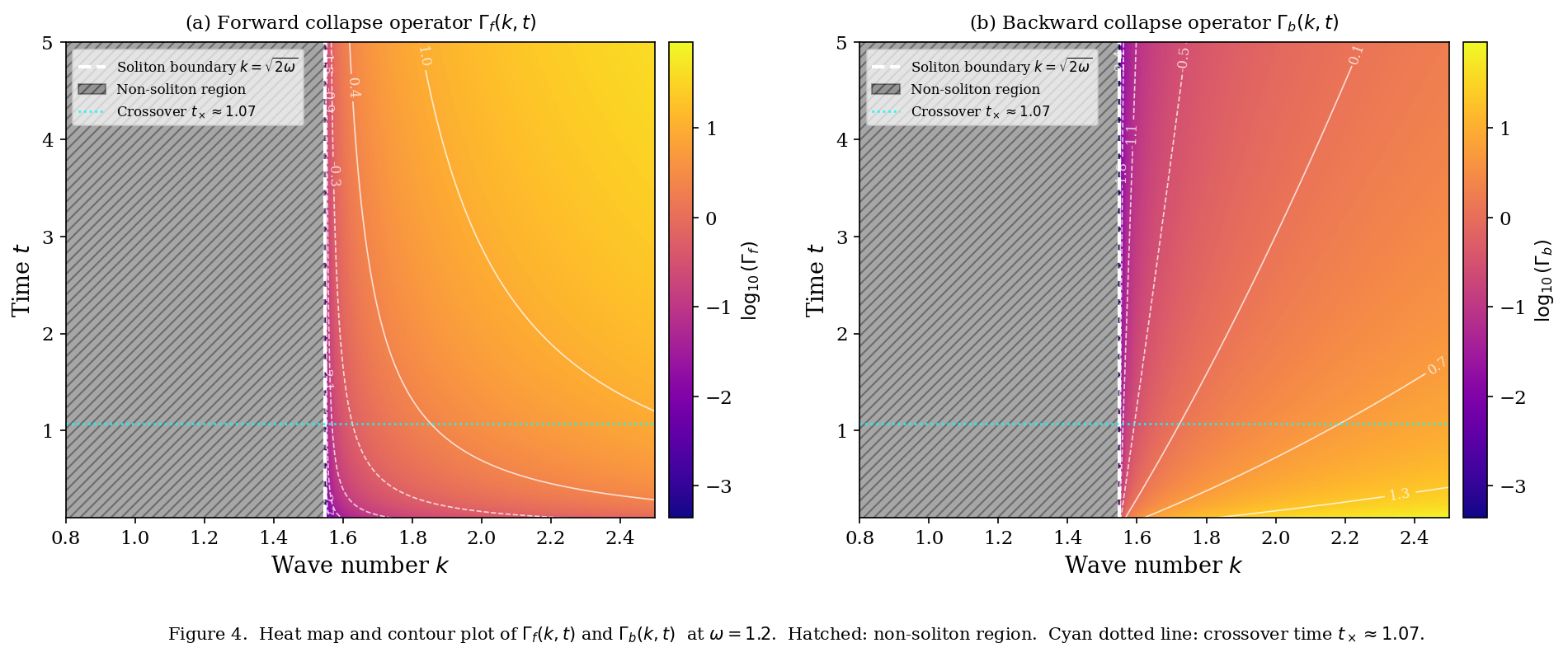}
  \caption{Heat map and contour plot of the forward collapse operator
           $\Gamma_f(k,t)$ (panel a) and backward collapse operator
           $\Gamma_b(k,t)$ (panel b) in the $(k,t)$ plane at $\omega=1.2$.
           Colour encodes $\log_{10}(\Gamma)$ on a shared scale; white
           contour lines label equal-value levels.  The hatched region
           ($k < \sqrt{2\omega}\approx 1.55$) is the non-soliton domain.
           The cyan dotted horizontal line marks the crossover time
           $t_\times\approx 1.07$ where $\Gamma_f = \Gamma_b$.
           Above $t_\times$, $\Gamma_f > \Gamma_b$ (forward dominates);
           below $t_\times$, $\Gamma_b > \Gamma_f$ (backward dominates).}
  \label{fig:coll_t}
\end{figure}

Figure~\ref{fig:heatmap} displays the complete $(k,\omega)$ parameter space
landscape of both collapse operators simultaneously as a heat map with
contour overlay, while Figure~\ref{fig:coll_t} shows their temporal evolution.
The heat map reveals the full two dimensional structure of the collapse
dynamics in a way that one dimensional slices cannot: the soliton boundary
$\omega = k^2/2$ appears as a sharp transition from the hatched non-soliton
region (upper left) to the active collapse region (lower right), and the
colour gradient immediately shows that $\Gamma_f$ (panel a) attains higher
values across the entire soliton-allowed region than $\Gamma_b$ (panel b),
confirming the forward--backward asymmetry in a single visual comparison.
Both panels share the same colour scale: $\Gamma_f$ reaches
$\log_{10}(\Gamma_f) \approx +1$ in the high-$k$, low-$\omega$ corner,
while $\Gamma_b$ reaches only $\log_{10}(\Gamma_b) \approx 0$ under the
same conditions, a tenfold difference in the collapse operator
magnitude visible directly from the colour contrast between the two panels.
The contour lines show that equal-$\Gamma$ surfaces run roughly parallel to
the soliton boundary, indicating that the proximity to the soliton cutoff
is the dominant factor controlling the collapse operator magnitude.
with the corresponding collapse operator asymmetry parameter shown in
Figures~\ref{fig:asym}.

All computed values of $\Gamma_f$ and $\Gamma_b$ are \emph{positive} across the entire
parameter space, consistent with the analytic form $\Gamma = \eta|\varPsi|^2(\hbar\omega)^2$
which is a product of intrinsically positive quantities.  This is consistent with
standard Continuous Spontaneous Localisation (CSL) theory, where the collapse operator
$\hat{L} = \int d^3x\, f(\mathbf{x}-\hat{\mathbf{x}})$ also gives $\Gamma > 0$
\citep{Bassi2013}.  The distinguishing feature of the present framework is not the sign
of $\Gamma$ but its \emph{energy modulation}: unlike the spatially uniform Gaussian
localisation function of CSL, the collapse operator here is amplified by the local
probability density $|\varPsi|^2$ and the excitation energy squared $(\hbar\omega)^2$,
concentrating collapse in high-density, high-excitation regions.  This is physically
analogous to the self-trapping transition in BEC double wells, where high local density
drives rapid nonlinear localisation \citep{Albiez2005, Salasnich2024collapse}.

Figure~\ref{fig:coll_t} shows the full $(k,t)$ landscape of both collapse operators
at $\omega=1.2$ as a heat map with contour overlay.  The two panels immediately reveal
the temporal asymmetry of the dynamics.  In panel~(a), $\Gamma_f(k,t)$ grows with
increasing $t$: the colour brightens from deep purple at small $t$ toward yellow to orange
at large $t$, confirming monotonic growth of the forward collapse operator with time.
In panel~(b), $\Gamma_b(k,t)$ shows the opposite behaviour: colour brightens with
decreasing $t$, confirming that the backward collapse operator is strongest at early
times and decays as $t$ increases.  The two panels share the same colour scale, so the
colour contrast between them at any fixed $(k,t)$ point directly encodes the
forward--backward asymmetry.

The cyan dotted horizontal line at $t_\times\approx 1.07$ marks the crossover time
where $\Gamma_f = \Gamma_b$.  Below this line panel~(a) is darker than panel~(b)
(backward dominates); above it panel~(a) is brighter (forward dominates).  The
crossover is visible as a horizontal colour inversion between the two panels and
constitutes a directly observable feature of the heat map.  At
$(k=2.0,\,\omega=1.0,\,t=1.0)$, the ratio $|\Gamma_f/\Gamma_b|\approx 1030$,
representing a strong quantitative prediction that distinguishes the present
framework from symmetric collapse models.

\subsection{Collapse Operator Asymmetry Parameter}

\begin{figure}[H]
  \centering
  \includegraphics[width=0.98\linewidth]{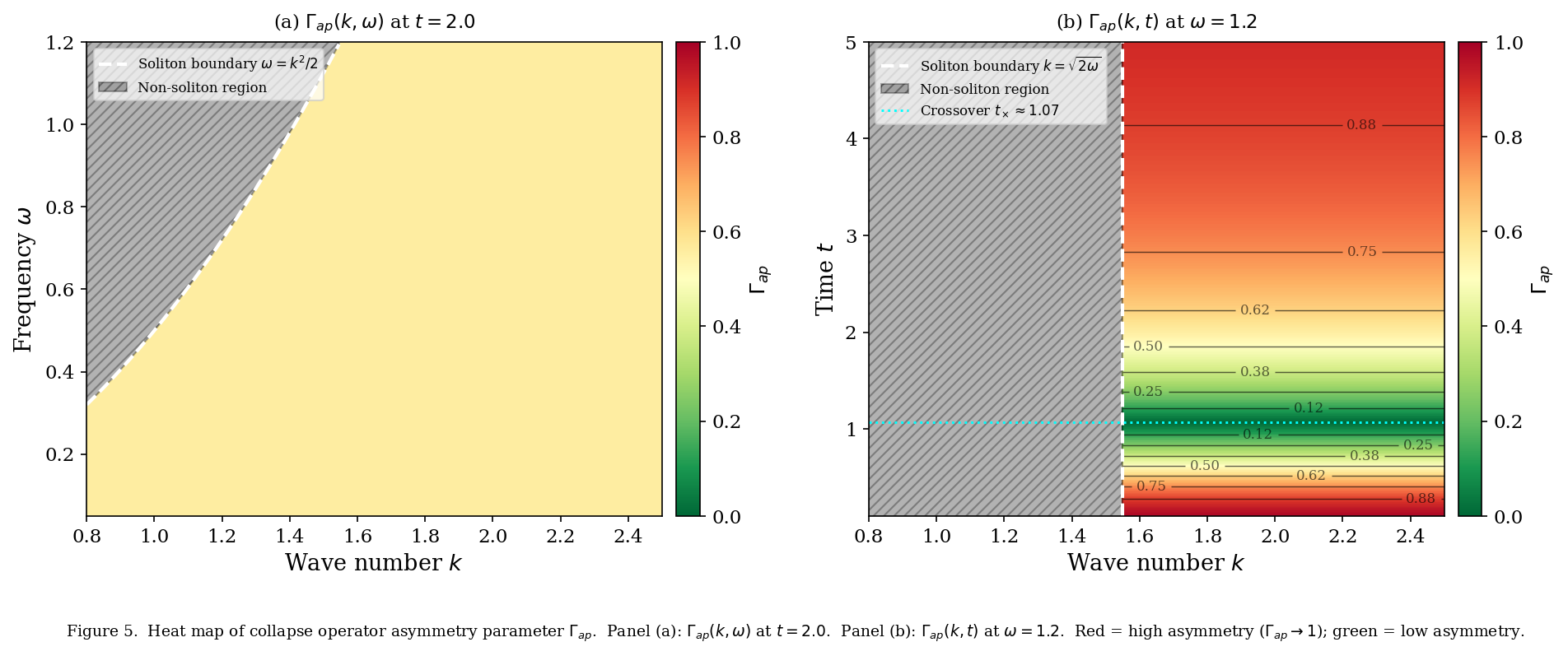}
  \caption{Two-panel heat map of the collapse operator asymmetry parameter
           $\Gamma_{ap} = |\Gamma_f-\Gamma_b|/(|\Gamma_f|+|\Gamma_b|)\in[0,1]$.
           Panel~(a): $\Gamma_{ap}(k,\omega)$ at $t=2.0$;
           panel~(b): $\Gamma_{ap}(k,t)$ at $\omega=1.2$.
           Colour: red = $\Gamma_{ap}\to1$ (strong asymmetry);
           green = $\Gamma_{ap}\to0$ (symmetric).
           Hatched: non-soliton domain.
           Cyan dotted line in panel~(b): crossover $t_\times\approx1.07$.}
  \label{fig:asym}
\end{figure}

Figure~\ref{fig:asym} presents the collapse operator asymmetry parameter
$\Gamma_{ap} = |\Gamma_f-\Gamma_b|/(|\Gamma_f|+|\Gamma_b|)$, bounded in $[0,1]$,
as a two-panel heat map: panel~(a) in the $(k,\omega)$ plane at $t=2.0$,
and panel~(b) in the $(k,t)$ plane at $\omega=1.2$.
The two panels together replace the three separate line plots previously used,
offering a richer and more complete picture of the asymmetry landscape.

Panel~(a) shows that $\Gamma_{ap}(k,\omega)$ is nearly uniform across the entire
soliton-allowed region, with values close to $0.67$ throughout.  The colour is
predominantly yellow, with only a slight gradient near the soliton boundary
$\omega = k^2/2$ (dashed white line) where the operators vanish and the asymmetry
parameter becomes indeterminate.  This near constancy of $\Gamma_{ap}$ in the
$(k,\omega)$ plane confirms that the structural asymmetry between the forward and
backward collapse operators is a universal property of the framework, independent
of the specific wave number or frequency chosen within the soliton-allowed region.
This is consistent with the analytic coupling asymmetry $\gamma_{ap} = 2/3 \approx 0.667$
derived in Section~\ref{sec:reduction}.

Panel~(b) shows a strikingly different picture in the $(k,t)$ plane.
Below the crossover time $t_\times\approx 1.07$ (cyan dotted line), $\Gamma_{ap}$
is low (green) — the forward and backward operators are comparable in magnitude and
the asymmetry is weak.  Above $t_\times$, $\Gamma_{ap}$ grows monotonically with $t$,
deepening from yellow to green through orange to deep red at large $t$, indicating that
the time-asymmetry strengthens continuously as the system evolves.
At $t=5.0$, $\Gamma_{ap}$ approaches $0.88$--$0.95$ across the soliton-allowed region,
confirming near total dominance of the forward over the backward collapse dynamics
at late times.

The contrast between the two panels encodes a central result of the present framework:
the \emph{structural} asymmetry (panel a) is fixed and parameter independent, while
the \emph{dynamical} asymmetry (panel b) grows in time, becoming more pronounced as
the system evolves away from the crossover point.  This combination of fixed structural
asymmetry and growing dynamical asymmetry is a distinctive prediction of the
energy-modulated time-asymmetric spontaneous collapse framework that is absent from
conventional symmetric collapse models.

\section{Conclusions}
\label{sec:conclusions}

We have developed the energy-modulated time-asymmetric spontaneous collapse
framework, deriving quantum symmetry breaking and irreversibility from stochastic
It\^{o} field reversal within a CQ-NLSE formalism.
Three physically necessary considerations \citep{Bassi2003} uniquely fix the
structure of any valid collapse model to be simultaneously nonlinear and stochastic.
From these considerations, an energy-driven collapse operator
$\Gamma = \eta|\varPsi|^2(\hbar\omega)^2$, proportional to the product of noise
strength, local probability density, and excitation energy squared, naturally
emerges from the physical operator choices and the It\^{o} isometry, specifically a product
of positive definite quantities confirming that collapse is enhanced by greater
amplitude, noise, and excitation energy.

The backward It\^{o} SDE, obtained by reversing the stochastic term and
incorporating the functional derivative correction, acquires a quintic coupling
five times that of the forward branch, yielding the universal asymmetry-coupling
parameter $\gamma_{ap} = 2/3$.  The CQ-NLSE possesses exact bright soliton
solutions in the soliton-allowed region, with forward and backward amplitudes
satisfying $A_f/A_b\approx 1.870$ and collapse-coupling parameters satisfying
$\gamma_b = 5\gamma_f$, confirming the structural asymmetry at the level of the
soliton solutions themselves.

Under the mean-field approximation valid for weak noise and large $n$, parametric
simulations and heat map analysis confirm the analytic predictions.  The $(k,t)$
heat map reveals the crossover time $t_\times\approx 1.07$ at which
$\Gamma_f = \Gamma_b$; above this time the forward branch dominates with
$|\Gamma_f/\Gamma_b|\approx 1030$ at $(k=2.0,\,\omega=1.0,\,t=1.0)$, constituting
a robust quantitative signature of time asymmetry that sharply distinguishes
the present framework from symmetric collapse models.

\noindent Sequel to this paper, our research directions include:
(i)~the development of a unified dynamical model embedded with collective
time-asymmetry diagnostics and collapse signatures, extending the present
energy-modulated framework toward a complete description of spontaneous
localisation in many body quantum systems;
(ii)~coupling the model to a realistic quantum measurement framework for
pointer state selection, thereby establishing a direct link between the
collapse operator $\Gamma = \eta|\varPsi|^2(\hbar\omega)^2$ and observable
measurement outcomes; and
(iii)~direct experimental tests of the asymmetry parameter $\gamma_{ap} = 2/3$
and the forward to backward collapse operator ratio $|\Gamma_f/\Gamma_b| \approx 10^3$
in ultracold $^7$Li bright soliton experiments \citep{Altamura2025, Carlesso2022natphys},
where the tunability of the scattering length and trap geometry provides natural
control over the nonlinear excitation frequency $\omega$ that governs the collapse operator.

\bibliographystyle{ieeetr}
\bibliography{refs}

\end{document}